%% file: paper.tex
\documentclass[a4paper]{jpconf}
\pdfoutput=1
\usepackage{graphicx}
\usepackage{todonotes}
\usepackage{hyperref}
\usepackage[hang]{subfigure}

\begin{document}
\title{Scaling MadMiner with a deployment on REANA}

\author{Irina Espejo$^1$, 
Sinclert Pérez$^2$, 
Kenyi Hurtado$^3$,
Lukas Heinrich$^4$ and 
Kyle Cranmer$^5$ }

\address{$^1$ New York University}
\address{$^2$ Shopify}
\address{$^3$ University of Notre Dame}
\address{$^4$ Technische Universität München}
\address{$^5$ University of Wisconsin--Madison}

\ead{iem244@nyu.edu}

\begin{abstract}
\input{Contents/Abstract}
\end{abstract}

\section{Introduction}
\input{Contents/Introduction}

\section{Workflow specification}
\input{Contents/Workflow-spec}

\section{Workflow deployment}
\input{Contents/Workflow-depl}

\section{Conclusion}
\input{Contents/Conclusion}

\ack
\input{Contents/thanks}

\pagebreak
\section*{References}
\bibliographystyle{plain}
\bibliography{BibTeX/bib-thesis-all}

\end{document}

%% file: Contents/Abstract.tex
MadMiner is a python package that implements a powerful family of multivariate inference techniques that leverage matrix element information and machine learning. This multivariate approach neither requires the reduction of high-dimensional data to summary statistics nor any simplifications to the underlying physics or detector response. In this paper, we address some of the challenges arising from deploying MadMiner in a real-scale HEP analysis with the goal of offering a new tool in HEP that is easily accessible.

The proposed approach encapsulates a typical MadMiner pipeline into a parametrized yadage workflow described in YAML files. The general workflow is split into two yadage sub-workflows, one dealing with the physics simulations and the other with the ML inference. After that, the workflow is deployed using REANA, a reproducible research data analysis platform that takes care of flexibility, scalability, reusability, and reproducibility features. To test the performance of our method, we performed scaling experiments for a MadMiner workflow on the National Energy Research Scientific Computer (NERSC) cluster with an HT-Condor back-end. All the stages of the physics sub-workflow had a linear dependency between resources or wall time and the number of events generated. This trend has allowed us to run a typical MadMiner workflow, consisting of 11M events, in 5 hours compared to days in the original study.

%% file: Contents/Introduction.tex
Searches for New Physics at the LHC require the use of complex event simulators in order to get precision measurements over subtle signals. In recent years, several Machine Learning (ML) applications in High Energy Physics (HEP) have shown promise of increased precision and scalability over traditional methods \cite{hepmlcommunityLivingReviewMachine}. A notable example is MadMiner \cite{brehmerMiningGoldImplicit2020}, a family of simulation-based inference techniques. These techniques bypass traditional dimensionality reduction by using ML to construct a surrogate for the intractable likelihood, thereby increasing precision. Hence, it would be beneficial to expand the application of MadMiner to a broader range of analyses.
Besides, the LHC undergoes future upgrades, and the deployment of MadMiner at scale becomes increasingly important. 

However, there are two main challenges for non-expert users to get started with MadMiner: first, the MadMiner pipeline generates simulator data on the fly with complicated software dependencies, and second, the pipeline involves several steps, making progress difficult for non-experts.

To address these challenges, we propose using REANA \cite{simkoREANASystemReusable2019}, a research data analysis platform that is containerized by nature and thus solves the issue of software dependencies. Additionally, REANA makes analysis reproducible by design, which is crucial in the context of ML in HEP, where reproducibility and plug-and-play compatibility with other ML tools are necessary due to the modular nature of simulation pipelines. Here we present a deployment of MadMiner using REANA with the following properties:
 \begin{itemize}
    \item It has a \textbf{shallow learning curve} after reading the MadMiner literature.
    \item It is \textbf{modular} so that each step or sub-workflow can be run independently.
    \item It has \textbf{interactive} access to output data and plots.
    \item It is \textbf{parameterized}, that is, the user can tweak inputs (parameters and files) without dealing with source code.
    \item It is \textbf{reproducible} and \textbf{reusable}.
    \item It \textbf{scales} with respect to the generation of events in the steps that involve physics simulators.
\end{itemize}

\subsection{MadMiner}
MadMiner \cite{brehmerEffectiveLHCMeasurements2019, brehmerGuideConstrainingEffective2018b, brehmerMadMinerMachineLearningBased2020c} is a python package that implements a powerful family of simulation-based inference techniques for High Energy Physics analysis. Two types of analysis are ubiquitous at the LHC: measurements of a physical variable and searches for New Physics. Both analyses require precision when analyzing high-dimensional data generated by the experiment and complex simulators.
These complex simulators produce a computationally intractable (marginal) likelihood both for maximum likelihood estimation (measurement) and hypothesis testing (searches).
There are different traditional approaches to this issue \cite{diggleMonteCarloMethods1984, rubinBayesianlyJustifiableRelevant1984b}, but all involve dimensionality reduction of the data to a low-dimensional space. The training data consist of an augmented dataset extracted from the latent states of the simulator. MadMiner automates all the steps necessary to apply simulation-based inference techniques and unites them in a pipeline.

\subsection{REANA}
REANA \cite{simkoREANASystemReusable2019} is a research data analysis platform built so that pipelines can be systematically reproduced and reused. Researchers encapsulate a research analysis by writing declarative yaml-based files. Those files describe a DAG workflow and the computations to execute at every step. To date, there is no standard approach to the reproducibility and sharing of ML models and pipelines in HEP, making REANA a valuable tool in this regard. REANA consists of a REANA client and a REANA cluster. The client is the front-end utility that offers the user an easy-to-use command line interface. The user loads the yaml files and selects a back-end. Automatically, REANA launches jobs in the back end, runs the workflow, and stores output files. The cluster component deals with the distribution of remote jobs. Different micro-services manage container-based computations submitted by the client.  Supported back-ends for REANA include Kubernetes, Slurm, and HT-Condor.

%% file: Contents/Workflow-spec.tex
The source code of the workflow can be found in Ref. \cite{madminer-workflow}.
The workflow topology is coded using \texttt{yadage} \cite{cranmerYadagePacktivityAnalysis2017}, a yaml-based workflow language.
Each stage of the workflow executes code written in \texttt{bash} and \texttt{Python} where the MadMiner functionality is provided by the  MadMiner package in Ref. \cite{madminer-repo}.
The workflow is divided into two parts: one dealing with Physics computations (in Ref. \cite{madminer-workflow-ph}), and the other one with Machine Learning computations (in Ref. \cite{madminer-workflow-ml}).
Containerization of dependencies was done using two Docker images, one containing Machine Learning dependencies (\texttt{PyTorch}) and another one containing the following physics dependencies: \texttt{MadGraph\_aMC v2.9.4}, \texttt{Pythia8}, \texttt{LHAPDF} and \texttt{Delphes 3}.
Each of the sub-workflows can be executed independently on a local computer using \texttt{yadage}. This way, users can re-run specific parts of a sub-workflow locally and study them more carefully. Additionally, the ML sub-workflow supports \texttt{MLFlow}, an ML platform offering multiple tools to keep track of ML metrics across experiment runs.

\subsection*{Physics sub-workflow}
Inputs to this sub-workflow are parameter information, benchmark information, observables, and cuts.
Users can change the physics signal by changing the cards located in this sub-workflow.
The following tasks, in order, form the DAG of the physics sub-workflow.
\begin{itemize}
    \item \textbf{Configure}: Load inputs and transform them into MadMiner objects.
    \item \textbf{Generate}: Generation in parallel of file configurations ready to be run by simulators.
    \item \textbf{Pythia}: Simulations in parallel of collisions with MadGraph and Pythia, as well as bookkeeping.
    \item \textbf{Delphes}: Lightweight simulation in parallel of collisions passing through the detector and bookkeeping.
    \item \textbf{Combine}: Partial combination of data files output by Delphes.
    \item \textbf{Multi-combine}: Complete combination of data files.
\end{itemize}

The following tasks, in order, form the DAG of the machine learning sub-workflow.
\subsection*{ML sub-workflow}
Inputs to this sub-workflow are the type of estimator of the likelihood ratio (parametrized or double-parametrized), sampling choices, and Neural Network architecture.
\begin{itemize}
    \item \textbf{Sampling}: Pre-processing of data from Delphes into training data with tasks like unweighting of events and data augmentation using the joint score.
    \item \textbf{Training}: Training of Multilayer Perceptron with input architecture to estimate likelihood ratio.
    \item \textbf{Evaluating}: Evaluation on test data of likelihood ratios or score estimator and limit setting.
    \item \textbf{Plotting}: Computation of diagnostic metrics and result plots using evaluation data.
\end{itemize}

The Physics workflow is structured as a map-reduce to enhance the scalability of the event generation step.
We observed that one map-reduce implementation could not go further than 1M events without a significant increase in wall time. That is why a partial and a complete \texttt{combine} step exists. The \texttt{multi-combine} step is the ``reduce" of a ``map" that is also a map-reduce. The number of times map-reduce is repeated in ``map" is an input by the user called \texttt{n\_procs\_per\_job}.

%% file: Contents/Workflow-depl.tex
\begin{figure}[!h]
    \centering
    \includegraphics[scale=0.7]{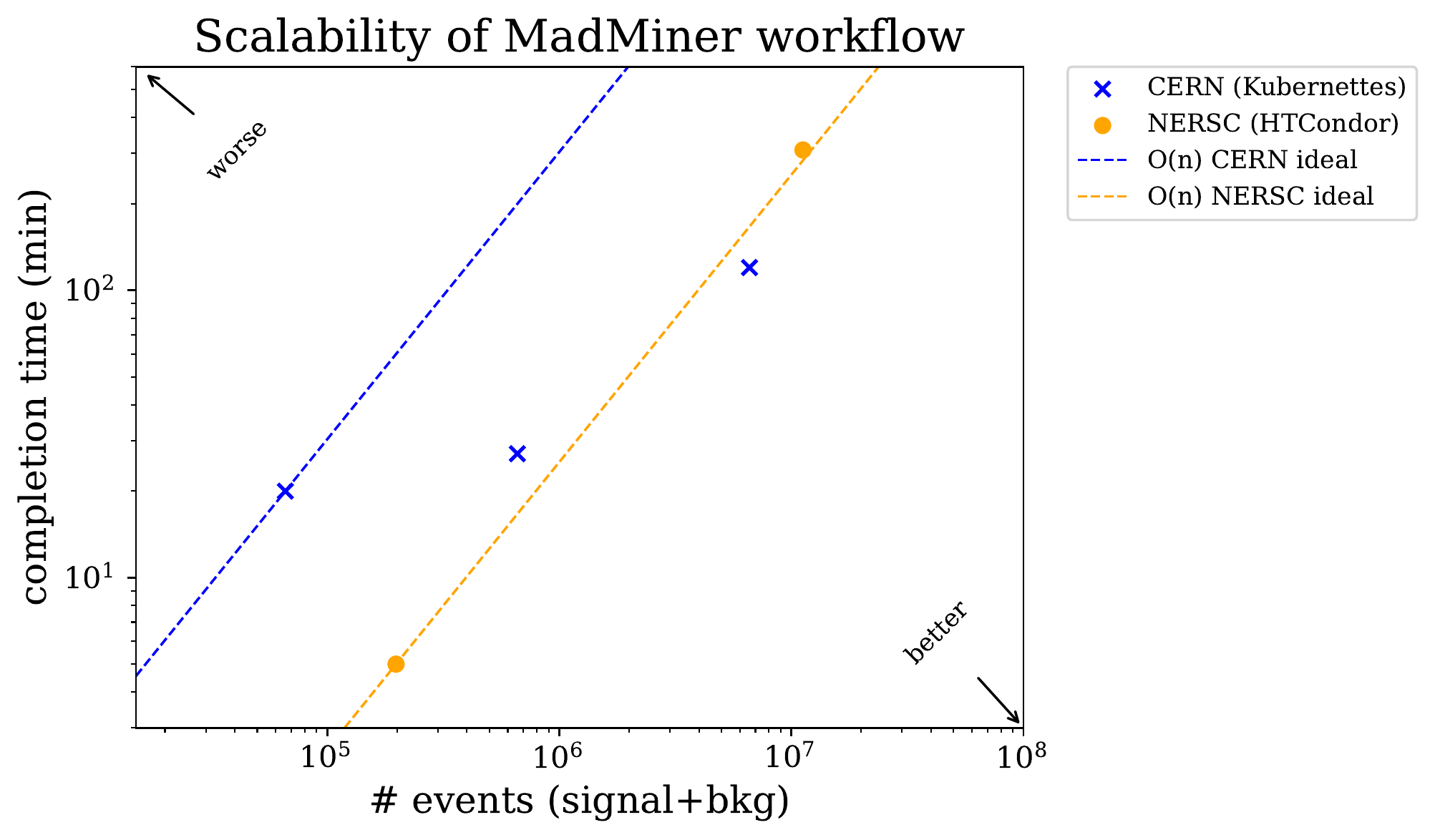}
    \caption{The plot compares the execution time of the MadMiner workflow as the number of Monte Carlo events generated increases. The study is carried out using two REANA cluster instances, one located at CERN that uses Kubernetes and the other located in NERSC that uses HT-Condor. Dashed lines represent ideal scalability; that is, the back end is able to distribute jobs efficiently among its resources as the number of jobs exponentially increases.}
    \label{fig:scale}
\end{figure}

The MadMiner workflow can be deployed locally and on a cloud-based system. Local deployment is controlled by \texttt{yadage}; this way, the researcher can run small-scale experiments with full control for debugging. Remote deployment uses REANA as the execution coordinator.
The researcher can execute the workflow either at a high level, with a \texttt{Makefile} provided in the source code, or the \texttt{reana-client} CLI. REANA provides an instance to monitor remote job deployment from the researcher to the local browser.
The workflow was successfully deployed at various HPCs, with the one deployed at the National Energy Research Scientific Computing Center (NERSC) being the most computationally intensive. This REANA instance is based on an HT-Condor back-end that generated 11 million simulation events in 5h 8 min. Out of 11M events, 1.1M were signal, and 10M were background before re-weighting. Compared to the original study,  Ref.   \cite{brehmerMiningGoldImplicit2020} generated 10 million events, and subsequent training spanned days. Out of 10M, 1.5M were signal, and the rest were background. All parameters, observables, cuts, and training configurations were similar in both cases. 
Regarding scalability, experiments were performed in two REANA instances, one at CERN using Kubernetes and the other instance at NERSC. In Fig. \ref{fig:scale}, the NERSC instance shows linear scaling for the total execution time with respect to an exponential increase in events generated up to a little over $10^7$ events. The CERN instance shows sub-optimal performance with respect to the ideal, indicating that a bottleneck in the distribution of resources for jobs was taking place. Both facilities had different resources, demand, and allocation policies, and at the time of the experiments at CERN in 2020-21, that instance was in development. The experiments we performed at CERN helped with the development, making later experiments at NERSC more robust.

\pagebreak
\begin{figure}[!h]
    \includegraphics[scale=0.08]{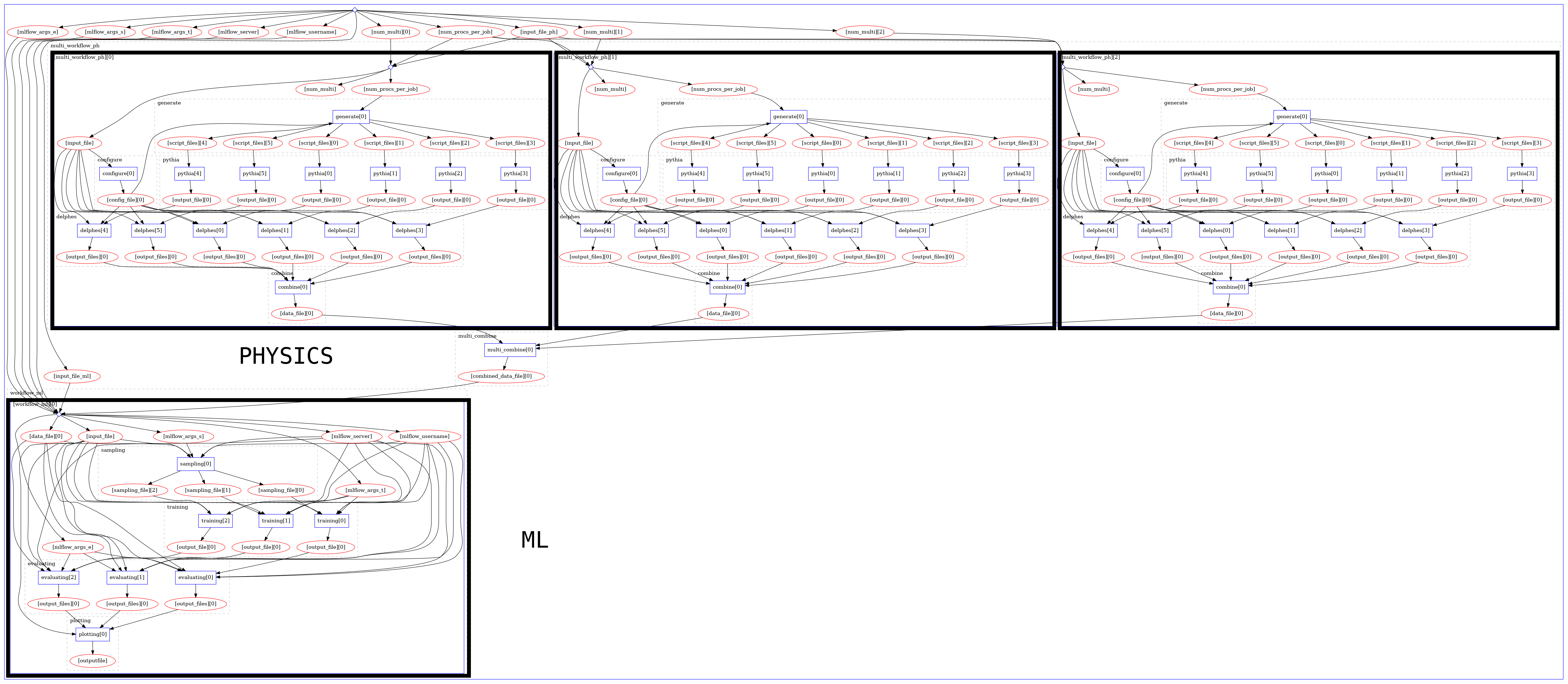}
    \caption{Diagram showing all the steps of a scaled-down workflow encapsulating a MadMiner pipeline. It starts with a configuration step that is followed by highly parallelizable simulation steps. Each of the upper black squares corresponds to a full mini-workflow that, in its turn, further parallelizes the \texttt{MadGraph}, \texttt{Pythia}, and \texttt{Delphes} steps.
    In a real-world experiment, this part would be replicated further to the right. After those, there is a combine step which is the principal bottleneck of the pipeline; this step collects all data from simulations and prepares it for training. The bottom black square corresponds to the ML sub-workflow where a Neural Network fits data coming from the physics sub-workflow. Finally, the workflow runs the inference and plotting steps that output results to the user.}
    \label{fig:workflow}
\end{figure}

%% file: Contents/Conclusion.tex
This paper presents a way of leveraging REANA's modular and scalable properties to run MadMiner pipelines. The deployment facilitates the use of machine learning to overcome likelihood intractability and increase precision in HEP searches. With this example, HEP researchers can now use MadMiner for their own analysis without the need for extensive coding or managing dependencies. This deployment is highly customizable, allowing users to change the number of events generated, observables and cuts, MLP architecture, and ways of approximating the likelihood with a neural network. Additionally, the deployment outputs are easily accessible on a REANA instance, and the deployment performs optimally in terms of scalability on the NERSC (HT-Condor) instance. The scalability experiments contributed to the development and benchmarking of the REANA cluster. In the future, it  would be useful to provide support for Slurm as a back-end, which is commonly used on HPCs and at CERN.

%% file: Contents/thanks.tex
\noindent This work was supported by the National Science Foundation under Cooperative Agreement OAC-1836650.
IE is supported by the National Science Foundation under NSF Award 1922658. SP, IE, and KC are supported by OAC-1841471.
KH is also supported by OAC-1841448. LH is supported by the Excellence Cluster ORIGINS, which is funded by the Deutsche Forschungsgemeinschaft (DFG, German Research Foundation) under Germany’s Excellence Strategy - EXC-2094-390783311.